\documentclass[10pt]{iopartmod}
\usepackage{amssymb}
\usepackage{amsmath}
\usepackage{makeidx}
\usepackage[dvips]{graphicx}

\input{tcilatex}

\begin{document}

\title{Giant Slow Wave Resonance for Light Amplification and Lasing}
\author{Alex Figotin and Ilya Vitebskiy}

\begin{abstract}
We apply the idea of giant slow wave resonance associated with a degenerate
photonic band edge to gain enhancement of active media. This approach allows
to dramatically reduce the size of slow wave resonator while improving its
performance as gain enhancer for light amplification and lasing. It also
allows to reduce the lasing threshold of the slow wave optical resonator by
at least an order of magnitude.
\end{abstract}

\maketitle

\section{Introduction}

Light-matter interactions play essential role in optics. Well-known effects
caused by light-matter interactions include: nonreciprocal circular
birefringence responsible for magnetic Faraday rotation, all kinds of
nonlinear effects, gain in active media, etc. At optical frequencies, the
above interactions are usually weak and can be further obscured by
absorption, radiation losses, etc. A\ way to enhance those interactions is
to place the optical material in a resonator. In cases of gain in active
media or Faraday rotation, a simple intuitive explanation for the resonance
enlacement of the respective effect invokes the idea that in a high-Q
resonator filled with the proper optical material, each individual photon
resides much longer compared to the same piece of material taken out of the
resonator. Since all the mentioned above interactions are independent of the
direction of light propagation, one can assume that their magnitude is
proportional to the photon residence time in the optical material. With
certain reservations, the above assumption does provide a hand-waving
explanation of the phenomenon of resonance enhancement. The above
qualitative picture may not directly apply to composite structures in which
the size of individual uniform components is comparable to or lesser than
the electromagnetic wavelength. At the same time, a number of popular
methods of resonance enhancement of light-matter interactions involve some
kind of composite dielectric structure, in which the typical feature size is
comparable to the electromagnetic wavelength.

The majority of high-Q optical resonators fall into one of the following two
categories: (a) cavity resonators, exemplified by optical microcavities in
which light confinement is produced by Bragg reflectors or by total internal
reflection (e.g., the whispering galery modes) and (b) slow wave resonators,
exemplified by periodic stacks of dielectric layers and photonic crystals
with two- and three-dimensional periodicity. Either kind of optical
resonator has certain advantages and disadvantages. Detailed information on
resonant microcavities can be found in numerous papers and monographs on
photonics (see, for example, \cite{RCavity 95} and references therein).
Further in this paper we will not discuss cavity resonators, focusing
instead on slow wave resonators and their applications in gain enhancement.
Gain enhancement is essential in light amplification and lasing. It can also
be used to offset losses produced by other material components of the
composite structure.

There can be two independent mechanisms of gain enhancement in a slow wave
photonic structure. The first one is associated with slow wave resonance,
which is often referred to as a Fabry-Perot resonance, or a transmission
band edge resonance. This mechanism works the best in high-Q slow wave
resonators where the amplitude of the oscillating electromagnetic field is
much higher than that of the incident light (for more information see, for
example, \cite{SWR 94,Yariv,Yeh,Chew,SL Scal1,SL Scal2,SL Joann,CR
Yariv04,OSU DBE}, and references therein). Another mechanism of gain
enhancement in a photonic structure is associated with the microscopic field
distribution inside the periodic dielectric array. Specifically, the
oscillating electric component of light should be concentrated in the gain
component of the composite structure, rather than in the other material
component. This latter idea is very similar to that used in \cite{PRB08} for
absorption suppression in magnetic photonic crystals.

The central idea of this paper is to use the so-called giant slow wave
resonance \cite{PRE05,PRA07} for gain enhancement. This approach allows to
reduce the size of slow wave resonator while dramatically improving its
performance as gain enhancer. It could be particularly attractive for light
amplification and lasing. In the latter case, not only it allows to reduce
the size of the optical resonator, but it can also significantly lower the
lasing threshold.

\subsection{Giant slow wave resonance}

Electromagnetic wave propagation in photonic crystals is qualitatively
different from any uniform substance. The differences are particularly
pronounced when the wavelength is comparable to the size $L$ of the unit 
\cite{Brill,LLEM,Joann,Yariv,Yeh,Chew,Notomi}. The effects of strong spatial
dispersion culminate at stationary points $\omega_{s}=\omega\left(
k_{s}\right) $ of the Bloch dispersion relation where the group velocity $%
u=\partial\omega/\partial k$ of at least one of the traveling Bloch wave
vanishes 
\begin{equation}
\frac{\partial\omega}{\partial k}=0,\ \text{at }k=k_{s},~\omega=\omega
_{s}=\omega\left( k_{s}\right) .   \label{SP}
\end{equation}
There are several reason for that. Firstly, vanishing group velocity always
implies a dramatic increase in density of modes at the respective frequency.
Secondly, vanishing group velocity implies certain qualitative changes in
the eigenmode structure, which can be accompanied by some spectacular
effects in light propagation. One such example is the frozen mode regime
associated with a dramatic amplitude enhancement of the wave transmitted to
the periodic medium \cite{PRB03,PRE03,PRE05B,JMMM06,WRM06,PRE06}. In this
paper, we focus on a different slow-wave effect, namely, on slow wave
resonance occurring in bounded photonic crystals at frequencies close to
photonic band edges. This slow wave phenomenon is also referred to as the
transmission band edge resonance. There are some similarities between the
frozen mode regime and the slow-wave resonance in plane-parallel photonic
crystals. Both effects are associated with vanishing group velocity at
stationary point (\ref{SP}) of the Bloch dispersion relation. As a
consequence, both effects are strongly dependent on specific type of
spectral singularity (\ref{SP}). A fundamental difference though is that the
frozen mode regime is not a resonance phenomenon in a sense that it is not
particularly sensitive to the shape and size of the photonic crystal. For
instance, the frozen mode regime can occur even in a semi-infinite periodic
structure, where the incident plane wave is converted to a frozen mode
slowly propagating through the periodic medium until it is absorbed \cite%
{PRB03,PRE03,PRE05B,JMMM06,WRM06,PRE06}. By contrast, in the case of a slow
wave resonance, the entire bounded periodic structure acts as a resonator,
resulting in strong sensitivity of the resonance behavior to the size and
shape of the photonic crystal.

The idea of giant slow-wave resonator was proposed in \cite{PRE05,PRA07}.
The giant transmission band-edge resonance is produced by slow waves
associated with a degenerate band edge (DBE). The DBE related slow wave
resonance is much more powerful compared to those achievable in common
layered and other periodic arrays displaying only regular photonic band
edges (RBE). Specifically, at the frequency of DBE related giant slow-wave
resonance, the electromagnetic energy density inside the photonic-crystal
can be estimated as%
\begin{equation}
\left\langle W_{DBE}\right\rangle \propto W_{I}N^{4},   \label{<W> DBE}
\end{equation}
where $W_{I}$ is the energy density of the incident wave and $N$ is the
total number of unit cells in the periodic stack. The respective Q-factor is%
\[
Q_{DBE}\propto N^{5}. 
\]
By comparison, the average EM energy density at a standard RBE related
resonance is%
\begin{equation}
\left\langle W_{RBE}\right\rangle \propto W_{I}N^{2}.   \label{<W> RBE}
\end{equation}
The respective Q-factor is%
\[
Q_{RBE}\propto N^{3}. 
\]
And this a huge difference! Our goal is to find out how to use this
advantage of the DBE-ralated giant slow wave resonance for light
amplification and lasing.

\section{Light propagation in periodic arrays of birefringent layers}

In this section we introduce some basic definitions and notations of
electrodynamics of stratified media composed of birefringent layers. Similar
results can be obtained in photonic crystals with two- or three-dimensional
periodicity, where the presence of birefringent optical materials is not
required. Also, one could consider corrugated optical waveguides.

\subsection{Transverse time-harmonic waves in stratified media}

In a stratified medium, the second rank tensors $\hat{\varepsilon}\left( 
\vec{r}\right) $ and $\hat{\mu}\left( \vec{r}\right) $ depend on a single
Cartesian coordinate $z$, normal to the layers. The Maxwell equations in
this case can be recast as%
\begin{equation}
\nabla\times\mathbf{\vec{E}}\left( \vec{r}\right) =i\frac{\omega}{c}\hat {\mu%
}\left( z\right) \mathbf{\vec{H}}\left( \vec{r}\right) ,\;\nabla \times%
\mathbf{\vec{H}}\left( \vec{r}\right) =-i\frac{\omega}{c}\hat{\varepsilon}%
\left( z\right) \mathbf{\vec{E}}\left( \vec{r}\right) .   \label{MEz}
\end{equation}
Solutions for Eq. (\ref{MEz}) are sought in the following form%
\begin{equation}
\mathbf{\vec{E}}\left( \vec{r}\right) =e^{i\left( k_{x}x+k_{y}y\right) }\vec{%
E}\left( z\right) ,\ \mathbf{\vec{H}}\left( \vec{r}\right) =e^{i\left(
k_{x}x+k_{y}y\right) }\vec{H}\left( z\right) .   \label{LEM}
\end{equation}
$\allowbreak$The substitution (\ref{LEM}) allows separation of the
tangential components of the fields into a closed system of four linear
ordinary differential equations%
\begin{equation}
\partial_{z}\Psi\left( z\right) =i\frac{\omega}{c}M\left( z\right)
\Psi\left( z\right) ,   \label{ME4}
\end{equation}
where%
\begin{equation}
\Psi\left( z\right) =\left[ 
\begin{array}{cccc}
E_{x}\left( z\right) & E_{y}\left( z\right) & H_{x} & H_{y}\left( z\right)%
\end{array}
\right] ^{T}   \label{Psi}
\end{equation}
and the $4\times4$ matrix $M\left( z\right) $ is referred to as the
(reduced) Maxwell operator. The normal field components $E_{z}$ and $H_{z}$
do not enter the reduced Maxwell equations (\ref{ME4}) and can be expressed
in terms of the tangential field components from Eq. (\ref{ME4}) as 
\begin{equation}
\begin{array}{c}
E_{z}=\left(
-n_{x}H_{y}+n_{y}H_{x}-\varepsilon_{xz}^{\ast}E_{x}-\varepsilon_{yz}^{%
\ast}E_{y}\right) \varepsilon_{zz}^{-1}, \\ 
H_{z}=\left(
n_{x}E_{y}-n_{y}E_{x}-\mu_{xz}^{\ast}H_{x}-\mu_{yz}^{\ast}H_{y}\right)
\mu_{zz}^{-1},%
\end{array}
\label{EzHz}
\end{equation}
where 
\[
n_{x}=ck_{x}/\omega,n_{y}=ck_{y}/\omega 
\]

The explicit expression for $M\left( z\right) $ in Eq. (\ref{ME4}) in \cite%
{PRE03}.

Waves propagating in the $z$ direction in transverse medium are referred to
as transverse waves 
\begin{equation}
\vec{E}\left( z\right) \perp z,\vec{H}\left( z\right) \perp z. 
\label{EH trsv}
\end{equation}
The respective Maxwell operator has the following simple form%
\begin{equation}
M\left( z\right) =\left[ 
\begin{array}{cccc}
0 & 0 & 0 & 1 \\ 
0 & 0 & -1 & 0 \\ 
-\varepsilon_{xy}^{\ast} & -\varepsilon_{yy} & 0 & 0 \\ 
\varepsilon_{xx} & \varepsilon_{xy} & 0 & 0%
\end{array}
\right] ,   \label{M trsv}
\end{equation}
where we put $\hat{\mu}=I$.

The values of $\Psi$ at any two different locations $z$ and $z^{\prime}$ are
related by the transfer matrix $T\left( z,z^{\prime}\right) $ defined by%
\begin{equation}
\Psi\left( z\right) =T\left( z,z^{\prime}\right) \Psi\left( z^{\prime
}\right) .   \label{T}
\end{equation}
The elements of the transfer matrix are expressed in terms of material
tensors and other physical characteristics of the stratified medium.

Let $\Psi_{I}$, $\Psi_{R}$, and $\Psi_{P}$ be the incident, reflected, and
transmitted waves, respectively. To the left of the stack (at $z<0$), the
electromagnetic field is a superposition of the incident and reflected
waves. To the right of the stack (at $z>D$), there is only the transmitted
wave. The field inside the periodic medium is denoted as $\Psi_{T}$. Since
all four transverse field components in (\ref{Psi}) are continuous functions
of $z$, we have the following boundary conditions at $z=0$ and $z=D$%
\begin{equation}
\Psi_{I}\left( 0\right) +\Psi_{R}\left( 0\right) =\Psi_{T}\left( 0\right) , 
\label{BC 0}
\end{equation}%
\begin{equation}
\Psi_{P}\left( D\right) =\Psi_{T}\left( D\right) .   \label{BC D}
\end{equation}

At any given frequency $\omega$, the time-harmonic field $\Psi_{T}\left(
z\right) $ inside the periodic stratified medium can be represented as a
superposition of Bloch eigenmodes, each of which satisfies the following
relation%
\begin{equation}
\Psi_{k}\left( z+L\right) =e^{ikL}\Psi_{k}\left( z\right) ,   \label{BW}
\end{equation}
or, equivalently,%
\begin{equation}
\Psi_{k}\left( z\right) =e^{ikz}\psi_{k}\left( z\right) ,\ \psi_{k}\left(
z+L\right) =\psi_{k}\left( z\right) .   \label{BM}
\end{equation}
The Bloch wave number $k$ is defined up to a multiple of $2\pi/L$. The
correspondence between $\omega$ and $k$ is referred to as the Bloch
dispersion relation. Real $k$ correspond to propagating (traveling) Bloch
modes. Propagating modes belong to different spectral branches $\omega\left(
k\right) $ separated by frequency gaps. In reciprocal and/or centrosymmetric
periodic structures, the Bloch dispersion relation is always symmetric with
respect to the points $k=0$ and $k=\pi/L$ of the Brillouin zone%
\begin{equation}
\omega\left( k_{0}+k\right) =\omega\left( k_{0}-k\right) , 
\label{w(k)=w(-k)}
\end{equation}
where%
\begin{equation}
k_{0}=0,\ \pi/L.   \label{k_0}
\end{equation}

\begin{figure}[tbph]
\scalebox{0.8}{\includegraphics[viewport=-30 0 500 500,clip]{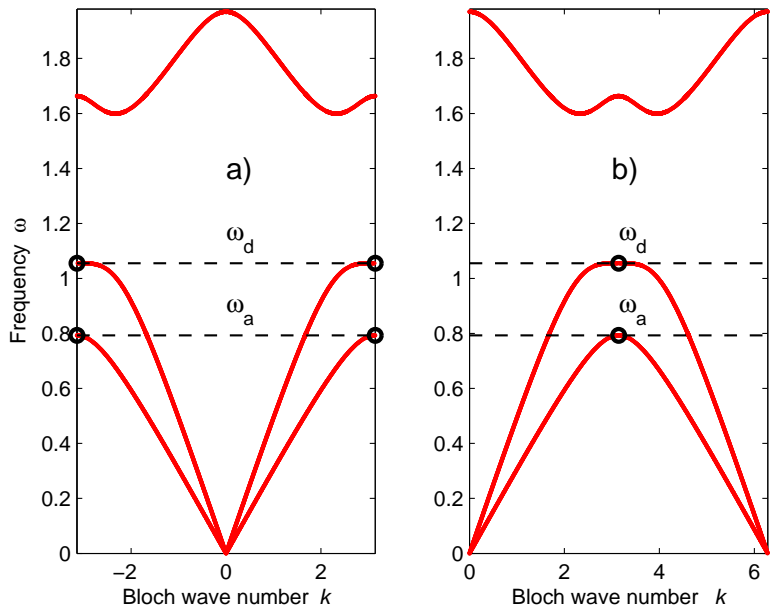}}
\caption{Bloch dispersion relation $k-%
\protect\omega $ with a degenerate photonic band edge at $\omega = \omega _{d}$.
 The Bloch wave number $k$ and the frequency $\omega$ are expressed in dimensionless units of $1/L$ and $c/L$.}
\label{dr2}
\end{figure}

In periodic structures composed of non-birefringent layers, every Bloch wave
is doubly degenerate with respect to polarization. If, on the other hand,
some of the layers display an in-plane anisotropy or gyrotropy, the
polarization degeneracy can be lifted. The respective $k-\omega$ diagrams
are shown in Fig. \ref{dr2}.

The speed of a traveling wave in a periodic medium is determined by the
group velocity $u=\partial\omega/\partial k$. Normally, every spectral
branch $\omega\left( k\right) $ develops stationary points (\ref{SP}) where
the group velocity of the corresponding propagating mode vanishes. Usually,
such points are located at the center and at the boundary of the Brillouin
zone%
\begin{equation}
k_{s}=k_{0}=0,~\pi/L.   \label{k_s=k_0}
\end{equation}
This is always the case in periodic layered structures composed of
non-birefringent layers, where all stationary points coincide with photonic
band edges. If, on the other hand, some of the layers in a unit cell are
birefringent, then in addition to (\ref{k_s=k_0}), some dispersion curves
can also develop a reciprocal pair of stationary points corresponding to a
general value of the Bloch wave number $k$, as shown in Fig. \ref{dr2}. The
respective portion of the $k-\omega$ diagram can be described as a split
band edge (SBE). The dispersion relation can develop a DBE or a SBE only if
the periodic layered array has birefringent layers with misaligned in-plane
anisotropy \cite{PRE05,PRE06}. Examples of such layered structures can be
found in \cite{PRE05,PRE06,PRA07}.

Unlike propagating modes, evanescent Bloch modes are characterized by
complex wave numbers $k=k^{\prime}+ik^{\prime\prime}$. Under normal
circumstances, evanescent modes decay exponentially with the distance from
the periodic structure boundaries. In such cases, the evanescent
contribution to $\Psi_{T}$ can be significant only in close proximity of the
surface or some other defects of the periodic structure. The situation can
change dramatically in the vicinity of a stationary point (\ref{SP}) of the
dispersion relation. At first sight, stationary points (\ref{SP}) relate
only to propagating Bloch modes. But in fact, in the vicinity of every
stationary point frequency $\omega_{s}$, the imaginary part $k^{\prime\prime}
$ of the Bloch wave number of at least one of the evanescent modes also
vanishes. As a consequence, the respective evanescent mode decays very
slowly, and its role may extend far beyond the photonic crystal boundary. In
addition, in some special cases, the electromagnetic field distribution in
the coexisting propagating and/or evanescent eigenmodes becomes very
similar, as $\omega$ approaches $\omega _{s}$. This can result in a
spectacular effect of coherent interference, such as the frozen mode regime 
\cite{PRB03,PRE03,WRM06,PRE06}. What exactly happens in the vicinity of a
particular stationary point essentially depends on its character and appears
to be very different for different types of singularity (\ref{SP}).

Using the transfer matrix (\ref{T}), the Bloch relation (\ref{BW}) can be
recast as%
\begin{equation}
T_{L}\Psi_{k}=e^{ikL}\Psi_{k}   \label{T BF}
\end{equation}
where $T_{L}$ is the transfer matrix of a unit cell%
\[
T_{L}=T\left( L,0\right) ,\ \Psi_{k}=\Psi_{k}\left( 0\right) . 
\]

At any given frequency, there are four Bloch eigenmodes, propagating and/or
evanescent, with different polarizations and wave numbers%
\begin{equation}
\Psi_{k1}\left( z\right) ,\Psi_{k2}\left( z\right) ,\Psi_{k3}\left( z\right)
,\Psi_{k4}\left( z\right) .   \label{4Psi k}
\end{equation}

In the presence of absorption and/or gain, all four Bloch wave numbers can
be complex at any finite frequency.

\subsection{EM energy flux in layered media}

The real-valued energy flux (the Poynting vector) associated with
time-harmonic field (\ref{Psi}) is%
\begin{equation}
S=\left[ \func{Re}\vec{E}\left( z\right) \times\func{Re}\vec{H}\left(
z\right) \right] =\frac{1}{4}\left(
E_{x}^{\ast}H_{y}-E_{y}^{\ast}H_{x}+E_{x}H_{y}^{\ast}-E_{x}H_{y}^{\ast}%
\right) .   \label{S(E,H)}
\end{equation}

Let $S_{I}$, $S_{R}$, and $S_{P}$ be the energy fluxes of the incident,
reflected, and transmitted waves, respectively. The transmission/reflection
coefficients $t$ and $r$ are defined as%
\begin{equation}
t=\frac{S_{P}}{S_{I}},\ r=-\frac{S_{R}}{S_{I}}.   \label{tn, rn}
\end{equation}

In case of lossy medium, we should also introduce the absorption coefficient 
$a$%
\begin{equation}
a=1-t-r.   \label{an}
\end{equation}

In case of active media, instead of absorption we should introduce the gain
coefficient $A$%
\begin{equation}
A=-1+t+r.   \label{An}
\end{equation}

In isotropic material with absorption and/or gain, the dielectric
permittivity is%
\[
\varepsilon=\varepsilon^{\prime}+i\varepsilon^{\prime\prime}. 
\]
Assuming that $\varepsilon^{\prime}>0$, in a lossy medium, the ratio $%
\varepsilon^{\prime\prime}/\varepsilon^{\prime}$ is positive 
\begin{equation}
g=\frac{\varepsilon^{\prime\prime}}{\varepsilon^{\prime}}>0,   \label{g}
\end{equation}
while in a gain medium, the same ratio is negative%
\begin{equation}
G=-\frac{\varepsilon^{\prime\prime}}{\varepsilon^{\prime}}>0.   \label{G}
\end{equation}

\section{Conclusion.}

In summary, let us reiterate the following basic questions. What happens to
the giant slow wave resonance in the presence of moderate losses or moderate
gain? How does the total gain $A$ defined in (\ref{An}) behave in the
vicinity of the resonance frequency. How do the above characteristics of the
giant slow wave resonance compare to those of the standard slow wave
resonance in the vicinity of a regular photonic band edge (RBE)?

Our analysis shows for small values of the ratio $G$ from (\ref{G}), the
total gain coefficient $A$ is proportional to $G$, both in the case of giant
slow wave resonance at DBE, and in the case of regular slow wave resonance
at RBE. The difference, though, is that in the former case%
\[
\text{For DBE resonance: }A\propto GW_{I}N^{5}, 
\]
while in the latter case%
\[
\text{For RBE resonance: }A\propto GW_{I}N^{3}, 
\]
where $W_{I}$ is the incident light intensity and $N$ is the number of
layers in the periodic stack. It means, for instance, that for $N=10$, the
giant slow wave resonance produces roughly $100$ times stronger gain,
compared to that of the regular slow wave resonance in the same stack. To
put it differently, to produce the same gain $A$, the stack with a DBE needs
by $N^{3/5}$ fewer layers, compared to the regular slow wave resonator. As
the ratio $G$ further increases, it reaches the threshold value, which is
the case of DBE-related slow wave resonator is lower by the factor $N^{2}$.
The above figures are rather impressive. They show that the giant slow wave
resonance is overwhelmingly better than the regular slow wave resonance,
when it comes to light amplification and lasing.

\textbf{Acknowledgment and Disclaimer:} Effort of A. Figotin and I.
Vitebskiy is sponsored by the Air Force Office of Scientific Research, Air
Force Materials Command, USAF, under grant number FA9550-04-1-0359.\ The
authors are thankful to A. Chabanov for stimulating discussions.

\pagebreak

\end{document}